\documentclass[11pt, preprint, showpacs]{revtex4}

\usepackage{graphicx}
\usepackage{amssymb}

\def\ba{\begin{array}}
\def\ea{\end{array}}
\def\be{\begin{equation}}
\def\ee{\end{equation}}
\def\ben{\begin{equation} \nonumber}
\def\een{\end{equation}}
\def\baray{\begin{eqnarray*}}
\def\earay{\end{eqnarray*}}

\begin{document}

\title{Cosmic Superstring Gravitational Lensing Phenomena: \\ Predictions for Networks of (p,q) Strings}
\author{Benjamin Shlaer$^1$}
\author{Mark Wyman$^{1,2}$}
\affiliation{$^1$ Laboratory for Elementary Particle Physics,
Cornell University,
Ithaca, NY 14853, USA \\
$^2$ Center for Radiophysics and Space Research, Cornell
University, Ithaca, NY 14853, USA}

\begin{abstract}
The unique, conical spacetime created by cosmic strings brings about distinctive gravitational lensing 
phenomena. The variety of these distinctive phenomena is increased when the strings have 
non-trivial mutual interactions. In particular, when strings bind and create junctions, rather than 
intercommute, the resulting configurations can lead to novel gravitational lensing patterns. 
In this brief note, we use exact solutions to characterize these phenomena,
 the detection of which would be strong evidence for the existence of complex cosmic string networks
 of the kind predicted by string theory-motivated cosmic string models.  We also correct some common
 errors in the lensing phenomenology of straight cosmic strings. 
\end{abstract}


\maketitle

\section{Introduction}
Some time after the existence of cosmic strings was proposed \cite{kibble}, 
several researchers recognized 
that the conical spacetime generated by cosmic strings leads to a unique gravitational 
lensing signature: undistorted double images \citep{lenshistory}; the discovery of 
even a single such gravitational lensing event would be seen as irrefutable
evidence for the existence of cosmic strings. 
Previous detailed studies of string lensing phenomena have focused on infinite
strings and string loops, whether straight or wiggly \citep{lensingstudy}. For standard, 
abelian Higgs strings, these are the only lensing effects that one would expect to find.

Recently, there has been a renaissance of interest in cosmic strings. This 
renewal was begun by the recognition that cosmic strings are copiously produced
in the aftermath of the brane collision that generates reheating in
 brane-world models of inflation in string theory \citep{branestrings}.
In addition to reviving interest in cosmic strings, these new studies have enriched cosmic
string phenomenology by proposing the 
existence of two basic cosmic string types: Fundamental, or F-strings, and one-dimensional
Dirichlet-brane strings, or D-strings. These two kinds of strings are able mutually to
interact to form bound states. These bound states are known as $(p,q)$ strings, as 
they are composed of $p$ F-strings and $q$ D-strings \citep{pqstrings}. String binding allows
for a variety of new observational phenomena, 
yet does not cause any cosmological catastrophes \citep{TWW05}. 
In particular, the existence of string binding interactions generically implies that there will be
 $Y$-shaped junctions of three strings that form each time there is a string binding event.
These $Y$-shaped junctions give rise to lensing phenomena that are qualitatively distinct 
from anything standard abelian Higgs models can produce.

Since cosmic strings are perhaps the only directly observable remnant of brane inflation, 
it is vital that we identify any distinctive effects that are peculiar to the cosmic ``superstrings" 
that are produced in brane-collision reheating.
Cosmic superstrings may be our best hope for directly observing some aspects of string theory. 
The observation of even a single distinctive lensing event -- one that 
could not be explained in an abelian Higgs model -- 
would be a ``smoking gun" for the existence of a non-trivial cosmic string network that
is the hallmark of cosmic superstring models. 
We note, however, that similar junctions are found in non-abelian string 
networks as well (e.g., the S3 networks studied in ref. \citep{penspergel, mcgraw}). 

\begin{figure}[h] 
   \centering
   \includegraphics[width=5in]{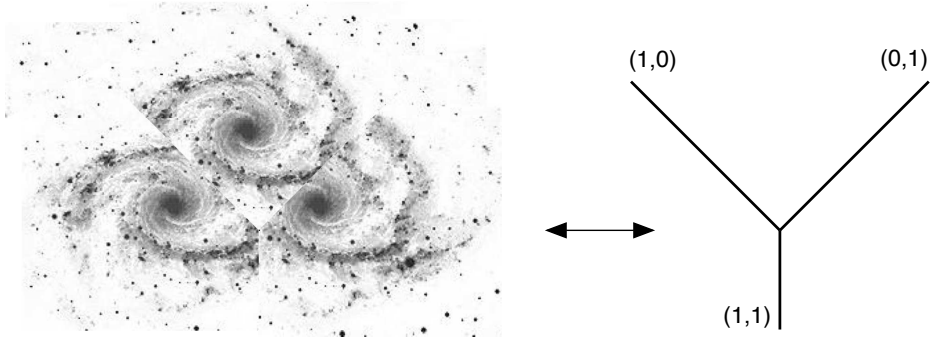} 
   \caption{Illustration of lensing of a single galaxy by a $(p,q)$ network junction.  Note that each 
   image is partially obscured, which is a generic feature of cosmic string lensing events \cite{obscured}.
   This image is an imaginative construction, not an actual observation. }
   \label{ynostrings}
\end{figure}

In this brief note, we describe the principal novel phenomena arising from the
binding junctions that characterize non-trivial networks of cosmic superstrings. In \S \ref{review},
we review lensing by a straight cosmic string, writing down some general formulae that have not
previously appeared in the literature. 
In Fig.\ref{ynostrings} we illustrate the quintessential
signature of a $(p,q)$ network junction. This is an imaginative construction of what 
the galaxy NGC 2997 \footnote{Anglo-Australian Observatory/David Malin Images} might 
look like if it were lensed by a $Y$-shaped string junction. In \S \ref{yjunctionsection}, we 
derive the simple procedure used to generate this image.

\section{Review of straight string lensing} 
\label{review}

\begin{figure}[h]
\centering
\includegraphics[width=1.5in]{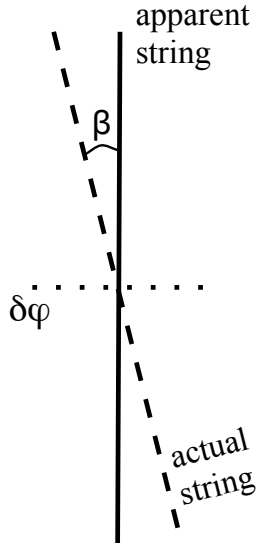}
\caption{A schematic view of the lensing due to a single string in the most general case; 
when the string's velocity is non-zero, the angular separation vector is perpendicular to the apparent
string, rather than to the actual string.} 
\label{definebeta}
\end{figure}

The lensing due to a straight string is surprisingly simple; this comes about because the surrounding
space-time is locally flat.  Many details of this lensing can nevertheless be quite subtle, and so here
we correct two errors found in the literature, namely the angular separation formula and the
orthogonality of the image pairs with respect to the observed cosmic string.  A string oriented in a direction  ${\bf \hat{s}}$ produces two images on opposite
sides of the string separated by an ``angular separation vector" $\vec{\delta\varphi}$ of
magnitude
\begin{equation}
|\vec{\delta\varphi}| = 8\pi G\mu \sqrt{\gamma^2 (1 + {\bf \hat{n}}\cdot{\bf v})^2 - \cos^2\theta}\frac{D_{s, cs}}{D_{s, o}}.
\label{lensformula}
\end{equation}
In our notation, bold face variables represent 3-vectors, bold hatted variables are unit 3-vectors, and over-arrows signify 2-vectors which
live in the plane orthogonal to the direction of sight.
 Here $G\mu$ is the dimensionless string tension, $\gamma = 1/\sqrt{1 - {\bf v}^2}$, ${\bf v}$ is
 the string velocity, ${\bf \hat{n}}$ is the unit vector directed along the line of sight, $\theta$ is
 the angle between the cosmic string and ${\bf \hat{n}}$ (i.e., it is defined by ${\bf \hat{n}}\cdot{\bf \hat{s}} = \cos\theta$), $D_{s, cs}$ is the distance from the source to the cosmic string, and $D_{s, o}$ is the distance from the
 source to the observer.
 Because cosmic strings are boost invariant along their axis, we will always work in the gauge where
  ${\bf v}$ satisfies $ {\bf v} \cdot {\bf \hat{s}} = 0$.
 Vilenkin \cite{lenshistory} pointed out that a straight cosmic string in motion will appear curved,
  like a large hyperbola in the sky; we note that, as expected,
   the apparent vanishing point of the hyperbola corresponds to the point where the magnitude
   of the angular separation vector goes to zero ($\vec{\delta\varphi} \rightarrow 0$). 
The reason for the apparent curvature is because what we see is the cosmic string world sheet intersected with our past light cone, 
 which we mentally project onto the $t = 0$ hyper-plane.  The cosmic string equations of motion
 are solved by 
 \begin{equation}
{\bf x}(\sigma, t) =  {\bf \hat{s}} \sigma/\gamma + {\bf v} t + {\bf b}
\end{equation}
with impact parameter  ${\bf b}$ (with respect to the origin/observer) orthogonal to both ${\bf \hat{s}}$ and ${\bf v}$.  The {\em observed} cosmic string
is described by the illusory embedding 
\begin{equation}
{\bf y}(\sigma, t) = {\bf x}(\sigma, t - |{\bf y}(\sigma, t)|)
\label{illusion}
\end{equation}
and so solving for ${\bf y}$ yields
\begin{equation}
{\bf y}(\sigma, t) = {\bf \hat{s}} \sigma/\gamma + {\bf v} ( \gamma^2t -  \sqrt{{\bf b}^2\gamma^2 + \sigma^2 + {\bf v}^2
\gamma^4 t^2}) + {\bf b}.
\end{equation} 
The apparent string then has orientation given by ${\bf \hat{s}}_{apparent} = {\bf y'}(\sigma, t)/|{\bf y'}|$, which is in general not equal to ${\bf \hat{s}}$.
These string orientation vectors may be pulled back onto the sky via
\begin{eqnarray}
\vec{s} &=& {\bf \hat{s}} - ({\bf \hat{n}}\cdot {\bf \hat{s}}){\bf \hat{n}}\\
\vec{s}_{apparent} &=& {\bf \hat{s}}_{apparent} - ({\bf \hat{n}}\cdot {\bf \hat{s}}_{apparent}){\bf \hat{n}}.
\end{eqnarray}
The difference between apparent and actual string orientation can then be characterized by the angle between
these two 2-vectors:
\begin{equation}
\cos\beta = \frac{\vec{s}\cdot\vec{s}_{apparent}}{|\vec{s}||\vec{s}_{apparent}|} = \frac{\cos\theta {\bf v}\cdot ({\bf \hat{n}}\times{\bf \hat{s}})\gamma}{\sin\theta\sqrt{\gamma^2 (1 + {\bf\hat{n}}\cdot{\bf v})^2 - \cos^2\theta}}.
\end{equation}
Interestingly, the relative angle between $\vec{\delta\varphi}$ and $\vec{s}$ can be shown
to be $\pi/2 + \beta$ which implies that the angular separation vector is always orthogonal
to the {\em apparent} cosmic string as in Fig. \ref{definebeta}.
\begin{equation}
\vec{\delta\varphi}\cdot\vec{s}_{apparent} = 0
\end{equation}

One might wonder how it is possible for a cosmic string that appears to be bent in
the shape of a hyperbola to produce an undistorted image.  In fact, the two images
are not completely identical:  They are moving at slightly different velocities, which gives
rise to both the Kaiser-Stebbins (blue-shift) effect, as well as a relative ``hyperbolic distortion" due to
the finite travel time of light.  So the distortion induced by an extremely relativistic string
is none other than the distortion that all moving bodies appear to have; c.f. Eqn. (\ref{illusion}).

\section{Lensing by $Y$-Junctions}
\label{yjunctionsection}
A feature of superstring networks is that they are composed of at least
two distinct string species: so-called fundamental or F-strings as well as D1-branes, 
or D-strings. These different types of strings can mutually interact via binding, creating
$(p,q)$ bound states composed of $p$ F-strings and $q$ D-strings \citep{pqstrings}.  The 
tension of such strings is given by 
\begin{equation}
\mu_{p,q} = \sqrt{p^2 \mu_F^2 + q^2 \mu_D^2}  
\end{equation}
 
Networks of such strings are cosmologically safe, as they are expected readily to go
to scaling \cite{TWW05}. 

\begin{figure}[h] 
   \centering
   \includegraphics[width=3in]{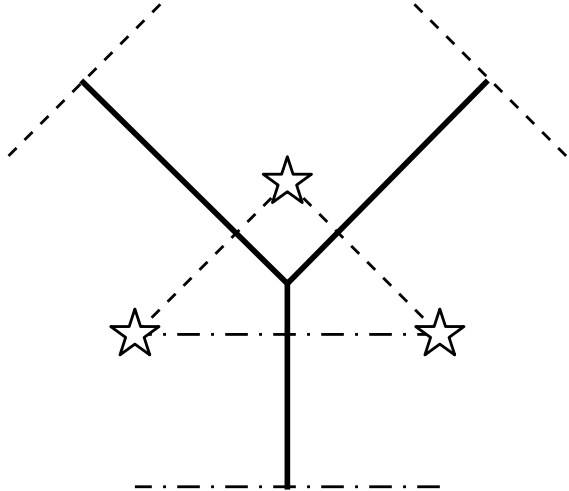} 
   \caption{The imaging pattern of a three-way junction. The dark lines represent the force-balanced
   string junction; in reality, strings themselves are invisible.
    The dotted and dot-dashed lines represent, 
   schematically, the angular separation vectors associated with the strings.  We suppress the
   heads of the vectors, since their orientation is arbitrary.
   For string tensions in the upper range
   allowed by observations -- that is, $G\mu \sim 10^{-7}$ \cite{gmulimits} -- the size of
   the angular separation vectors would be
   of order 1 arc-second. The stars represent the lensed images, with the angular separation vectors
   drawn in for illustrative purposes.}
   \label{yjunction}
\end{figure}

It is in the region near the $Y$-shaped junctions formed after collisions that the new 
string lensing effects are seen.  In Fig. \ref{ynostrings}, we showed an image as it might actually appear, with the invisible
strings and angular separation vectors suppressed.
In Fig. \ref{yjunction}, we show a mock-up of
the triple image formed in the vicinity of a $Y$-junction together
with the strings that produce the image. 
In brief, each leg of the $Y$-junction lenses exactly like an infinite straight string.

To motivate the above result, let us begin by considering a source simultaneously imaged by two
long strings.

\begin{figure}[h] 
   \centering
   \includegraphics[width=3in]{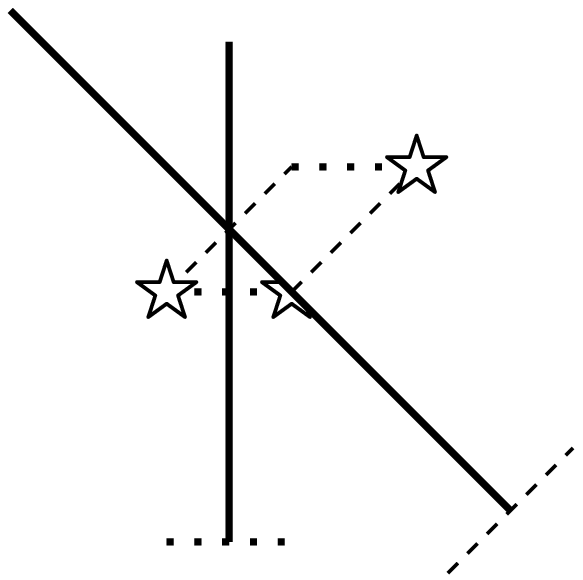} 
   \caption{The imaging pattern for two overlapping cosmic strings. The dotted and dot-dashed lines represent, schematically, the angular separation vector associated with the strings.  Notice that
   only the object and two of the
   three images are visible.}
   \label{2strings}
\end{figure}

For two overlapping strings, it is straightforward to construct a set of rules for the multiple imaging
of a single source.  Each straight cosmic string has an associated two dimensional ``angular separation vector" 
the length of which is given by Eqn. (1).  For string tensions
in the upper range allowed by observations \cite{gmulimits}, the magnitudes of these
 angular separation vectors are of order 1 arc-second. The 
orientation of the angular separation vector is perpendictular to the associated cosmic string. 

\begin{enumerate}
\item Begin with an original image (i.e. the object); the choice of which image to begin with is
arbitrary.  
\item Construct a parallelogram originating at the object and generated by the angular separation 
      vectors associated with each cosmic string; each corner represents a new image.
	\begin{enumerate}
            \item Each image (except the object) will be associated with the set of cosmic strings whose 
                  angular separation vectors were used to create it.
            \item If exactly those strings that are associated with an image  -- and no others  -- lie
                 between the object and that image, then
                  that image will be visible; otherwise, that image will be invisible.	
		\end{enumerate}
\item Thus, given the existence of one visible image (the object), the location and status -- visible or 
      invisible -- of the other three images 
      is known. An example of this is shown in Fig. \ref{2strings}.
\item This procedure is consistent with the fact that a visible image is made invisible when and only when any cosmic string moves across it, and an invisible image is made visible only when a cosmic string moves across it.
\end{enumerate}

\begin{figure}[h] 
   \centering
   \includegraphics[width=3in]{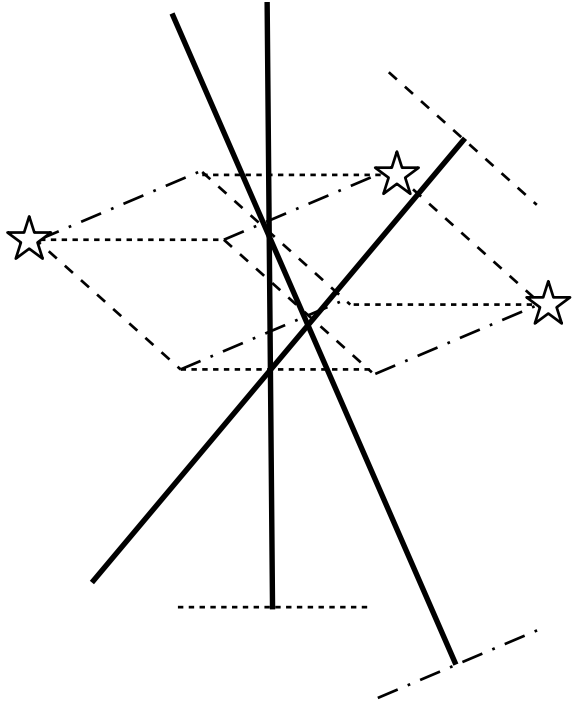} 
   \caption{The imaging pattern for three overlapping cosmic strings.  Only the object and two of
   the seven images are visible.}
   \label{3strings}
\end{figure}

For three overlapping strings, we follow the same procedure as before, with the same rules. The
only difference is that rather than forming a parallelogram, three angular separation vectors lead to 
a parallelepiped. We show an example of this sort of diagram in Fig. \ref{3strings}.

Finally, let us stipulate that the three strings are coplanar and intersect at one point.  
If they are to be in force balance, then their orientation unit vectors must obey
\be
\sum_{i = 1}^3\mu_i\gamma_i{\bf\hat{s}} = 0.
\label{force}
\ee 

The $\gamma$-factors correct for the Lorentz contraction caused by boosting in a direction not
perpendicular to all three strings.  

If the three infinite strings satisfy Eqn. (\ref{force}) then one may ``cut-and-paste" them into 
two $Y$-junctions without changing the energy-momentum tensor.  One of the $Y$-junctions
can then be pulled away to reveal a single junction and its lensed images.

When cosmic strings intersect at a three-way junction, the background metric is solvable if and 
only if \cite{Csaki:1999mz} the strings obey the force balance equation.
Dynamically, however, the force balance equation is always satisfied since
the neighborhood of the junction point can be taken to have an arbitrarily small
mass, thus allowing it to respond instantaneously to any net force.  
This would not be the case if the vertex had a large mass (e.g., if it were a monopole).  
Thus there is a consistency check that must be performed for lensing by $Y$-junctions:  each string
lenses the two images surrounding it, and so the picture is over determined.  In other words, 
each object has two images which must also satisfy the lensing equation of the string separating them.
This means we must check that the angular separation vector triangle closes:
\be
\sum_{i = 1}^3 \vec{\delta\varphi}_i = 0.
\label{close}
\ee
In fact, we find that the force balance Eqn. (\ref{force}) is a sufficient condition for Eqn. (\ref{close}) to be 
satisfied.  This could be inferred from the first claim of this paragraph, but is also easy to verify directly
using the equations found in \S \ref{review}.

\section{Conclusion}

We have presented an exact solution for both the lensing of three-string junctions
and for multiple overlapping strings.  We do not yet know how prevalent such junctions
would be in realistic networks of (p,q) strings. Since they are expected to be somewhat rare,
however, our chief hope for locating such a junction would be, first, to locate an actively lensing cosmic 
string. As yet, no astronomical observations have yielded unambiguous evidence for such a
lensing event, though there is currently a candidate observation, CSL-1 \citep{sazhin}, awaiting
confirmation or refutation by the Hubble Space Telescope \citep{hoganprivate}. In passing, 
we note that, since high velocities generally enhance lensing
 (see Eqn. \ref{lensformula}), there cannot be a firm lower bound placed on the tension
 of the inferred string tension from this event, if it proves to be authentic.
Should such an event be found, we might hope to track it across the sky by interpolation between
further lensing events or through the Kaiser-Stebbins effect
\cite{kaiserstebbins} until a junction could be located along its length. If it so happens
that CSL-1 is confirmed as an authentic cosmic string lensing event, an eager modern-style search would doubtless ensue in hopes of tracking
down further such events; in such a scenario, the sort of cross-sky tracking suggested
above might become possible. Within any such search, the discovery of even one triple imaging
event as described in this paper would be an unmistakable indicator of the existence of a 
cosmic string network with non-trivial interactions, the very kind predicted in brane inflationary 
models. Finding this sort of smoking gun -- so rare as to be invisible to serendipitous discovery --
might be possible in such a scenario, giving us hope for an experimental examination of
string theory unthinkable even a few years ago.

\acknowledgements
We thank Henry Tye and Ira Wasserman for illuminating discussions, Nick Jones 
and Mark Jackson for helpful comments (over wine on the banks of the Seine)
 and the Anglo-Australian 
Observatory / David Malin Images for allowing us to use their image of NGC 2997. 
M.W. acknowledges  support from the National Science Foundation, 
the National Aeronautics and Space Administration, and the Cornell Boochever Fellowship.

\appendix
\section{Fly's Eye Effect}

\begin{figure}[h] 
   \centering
   \includegraphics[width=3in]{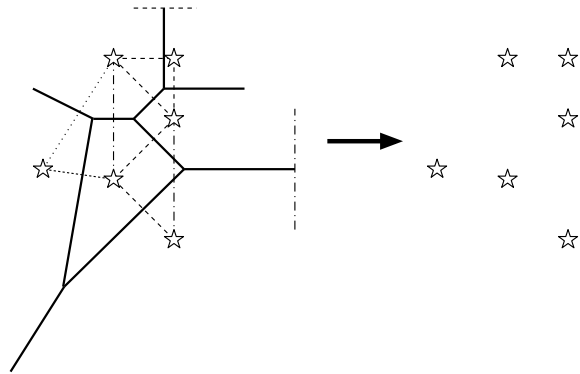} 
   \caption{This set-up of strings and associated images is unlikely ever to
   occur in nature. It is included for illustrative purposes.}
   \label{flyseye}
\end{figure}

A surprising secondary effect can be obtained if a large number of $Y$-junctions 
are arranged near one another: a single source can be identically imaged 
many, many times. We include this effect, though it is unlikely to arise in nature -- where the
junctions tend to speed apart relativistically -- as an example of the extreme limit of the
$Y$-lensing phenomenon. We further note that, while unlikely in present models, it has 
frequently been mentioned in the past \citep{pqstrings,penspergel,mcgraw} that networks of
strings that become frozen could have a fly's eye-style structure. We illustrate, on a small 
scale, what such a configuration could look like in Fig. \ref{flyseye}; 
on a cosmic scale, the number of links and associated images could, of course, 
be multiplied almost indefinitely.

\end{document}